\documentclass[journal=jacsat,manuscript=article]{achemso}

\usepackage[none]{hyphenat}

\usepackage[pdftex,colorlinks=true,linkcolor=Cerulean,citecolor=Cerulean,urlcolor=Cerulean]{hyperref}
\hypersetup{linkcolor=blue,citecolor=blue,urlcolor=blue}
\hypersetup
{
	pdfauthor={
		J. Enrique V\'azquez-Lozano (juavazlo@upvnet.upv.es),
		Jeremy J. Baumberg (jjb12@cam.ac.uk) \&
		Alejandro Mart\'inez (amartinez@ntc.upv.es)},
	pdfsubject={
		Nanophotonics Technology Center [Universitat Polit\`ecnica de Val\`encia] \& 
		NanoPhotonics Centre - Cavendish Laboratory - Department of Physics [University of Cambridge]},
	pdftitle={
		Enhanced excitation and readout of plasmonic cavity modes in NPoM via SiN waveguides for on-chip SERS},
	pdfkeywords={Surface-Enhanced Raman Scattering (SERS), Nanoparticle-on-a-Mirror (NPoM), integrated photonics, optical confinement and manipulation}
}

\usepackage{chemformula} 
\usepackage[T1]{fontenc} 

\author{J. Enrique V\'azquez-Lozano}
\affiliation{Nanophotonics Technology Center, Universitat Polit\`ecnica de Val\`encia, Camino~de~Vera~s/n, 46022 Valencia, Spain}

\author{Jeremy J. Baumberg}
\affiliation{NanoPhotonics Centre, Department of Physics, University~of~Cambridge,\\Cambridge CB3 0HE, United Kingdom}

\author{Alejandro Mart\'inez}
\affiliation{Nanophotonics Technology Center, Universitat Polit\`ecnica de Val\`encia, Camino~de~Vera~s/n, 46022 Valencia, Spain}
\email{amartinez@ntc.upv.es}

\title{Enhanced excitation and readout of plasmonic cavity modes in NPoM via SiN waveguides for on-chip SERS}

\abbreviations{NPoM,SERS,SiN,TM,NIR,NP,SAM}

\begin{document}
\maketitle
\sloppy
	
	
%
%
%
%
%
	

\begin{abstract}
\textbf{Metallic nanoparticle-on-a-mirror (NPoM) cavities enable extreme field confinement in sub-nm gaps, leading to unrivaled performance for nonlinear processes such as surface-enhanced Raman scattering (SERS). So far, prevailing experimental approaches based on NPoMs have been performed by means of free-space light excitation and collection under oblique incidence, since the fundamental radiatively-coupled NPoM mode does not scatter in the normal direction. Retaining this working principle, here we numerically show that plasmonic cavity modes in NPoM configurations can be efficiently excited in an integrated SERS approach through TM guided modes of silicon nitride (SiN) waveguides. Intensity enhancements beyond 10$^{5}$ can be achieved for gap spacings around 1 nm. So as to reduce unwanted SiN Raman background, the output Stokes signals are transferred to transversely placed waveguides, reaching coupling efficiencies of up to 10\%. Geometrical parameters such as the gap thickness as well as the radius and position of the nanoparticle provide full control over the main spectral features, thereby enabling us to engineer and drive the optical response of NPoMs for high-performance SERS in Si-based photonic integrated platforms.}
\end{abstract}


Plasmonic cavities allow for extreme subwavelength confinement of optical fields in nm-scale gaps between metallic surfaces \cite{Schuller2010,Zhu2016}. Such extreme localization enormously boosts nonlinear processes, including Raman scattering, which is usually termed \textit{surface-enhanced Raman scattering} (SERS) when plasmonic modes contribute to enhancing their response \cite{GarciaVidal1996,Kneipp1997,Nie1997,Kneipp2008,Ding2017}. Among the different implementations of plasmonic nanocavities, to date, the so-called \textit{nanoparticle-on-a-mirror} (NPoM) approach probably provides the deepest and most robust confinement, which takes place in the sub-nm gap separating the nanoparticle (NP) from an underlying metallic mirror \cite{Mock2008,Baumberg2019}. Indeed, by means of self-assembly fabrication techniques, even picocavities can be dynamically formed in the gap of a NPoM when exciting the fundamental radiative mode (elsewhere referred to as $(l=1,m=0)=(10)$, in connection with the terminology often used to deal with spherical harmonics) \cite{Ciraci2012,Kongsuwan2020}, thereby enabling ultra-low mode volumes $<1$ nm$^3$ \cite{Maier2006,Chikkaraddy2016,Benz2016a,Carnegie2018}. Thus, the realization of SERS in NPoM configurations becomes highly efficient because of the extreme confinement, facilitating also the direct observation of Raman modes which are not accessible under other plasmonic approaches \cite{Baumberg2019}.

Plasmonic cavities can also be integrated with all-dielectric photonic waveguides, giving rise to hybrid architectures of plasmonic-photonic integrated circuits \cite{Fortuno2016}. Within this scenario, plasmonic elements are compactly arranged on a chip to perform different functionalities, exhibiting low-power consumption and subwavelength footprint. This approach has already evidenced both the theoretical and experimental feasibility of on-chip SERS with single plasmonic nanocavities \cite{Peyskens2016,Dhakal2016a,Raza2019}, opening up unique prospects for the realization of multiplexed Raman spectroscopy in Si-based photonic platforms \cite{Peyskens2018,Losada2019}. Thus far, though, proposals for integrating plasmonic cavities into dielectric waveguides have made use of lithographically-defined nanoantennas with complex shapes (e.g., bowtie gold nanoantennas), so limiting the minimum size of the plasmonic gap to $\sim10$ nm \cite{Zhang2015}. This strongly limits the maximum field confinement attainable in the gap, and, consequently, the efficiency of nonlinear processes such as SERS.

In this work, we show from numerical simulations that the fundamental mode of NPoM plasmonic cavities can be efficiently excited by transverse magnetic (TM) guided modes of silicon nitride (SiN) strip waveguides. The in-plane on-chip driving of light bridges the problem of complex illumination schemes featured with free-space excitation \cite{Turk2019}, since the fundamental radiative NPoM mode, i.e., the $(10)$-mode, which dominates for small gap~sizes~($<5$ nm), does not scatter out along the normal to the mirror plane \cite{Kongsuwan2020,Christiansen2020}. Furthermore, from the Raman scattering standpoint, it is also shown that the enhanced Stokes fields generated by a point-like dipolar source in the sub-nm gap can be straightforwardly collected by dielectric waveguides \cite{Dhakal2014,Dhakal2015,Dhakal2016b}. In this regard, so as to mitigate the spurious Raman background noise originated from the input SiN waveguide \cite{Dhakal2017}, we use a specific ``T-shaped'' configuration. For the sake of robustness and comparison with earlier results, as well as with a realistic experimental setup (on account of deviations from the nominal values in the fabrication process) \cite{Lee2019}, we perform a sweep of the characteristic geometrical parameters such as the gap size, the NP radius and its relative position over the mirror plane. By varying these parameters one can gain control over the main spectral features, tuning the peak wavelength and the magnitude of both the intensity enhancement factor and the coupling strength.

\begin{figure}[t!]
	\centering
	\includegraphics[width=0.9\linewidth]{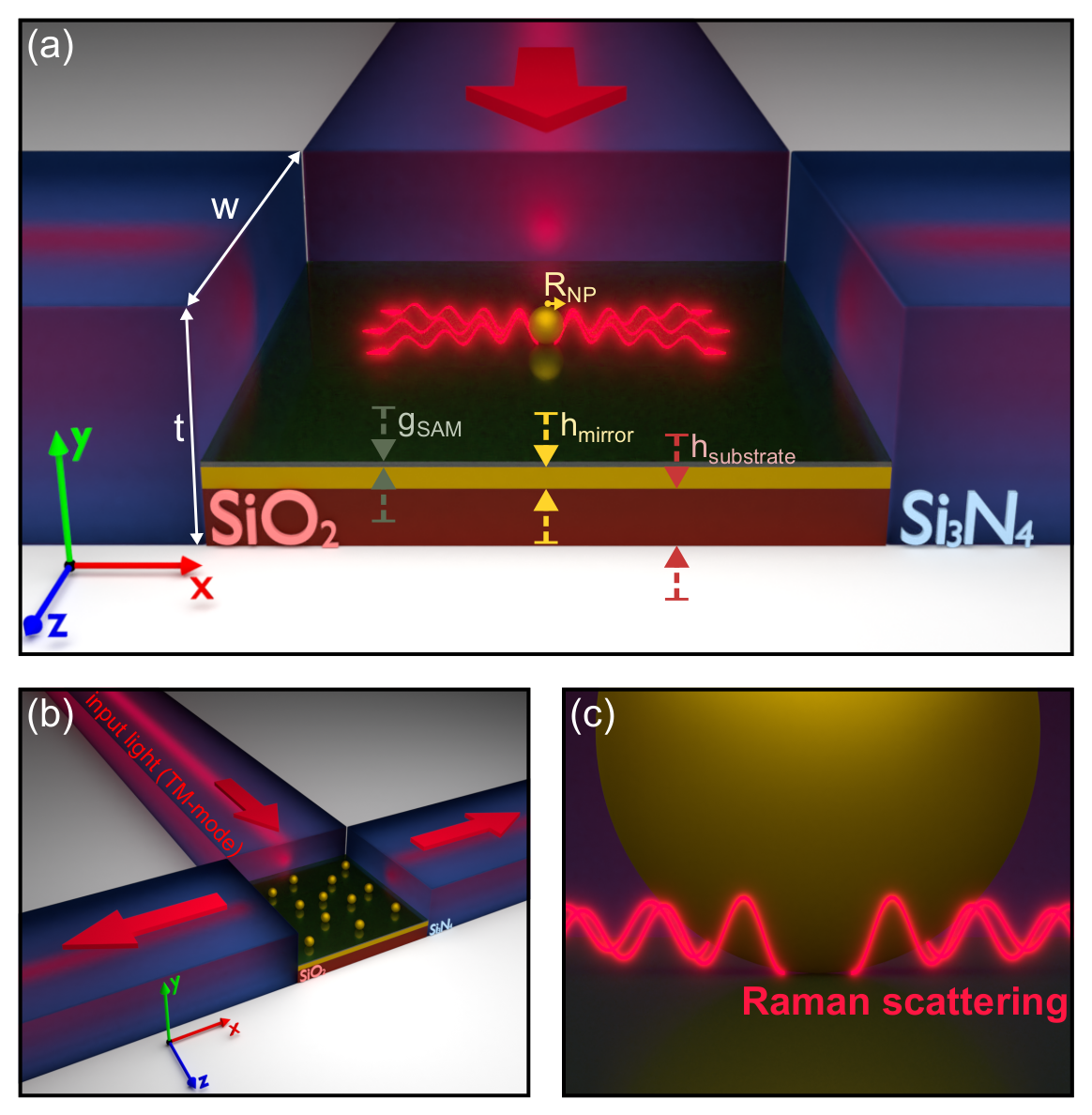}
	\caption{(a) Schematic depiction of the model used in the numerical simulations of NPoM for on-chip SERS. (b) Illustration of the proposed design in a realistic experimental setup. (c) Zoom-in view of the region where SERS process is to take place.}
	\label{fig01}
\end{figure}

The schematic of the proposed NPoM system for on-chip SERS is depicted in Figure \hyperref[fig01]{1}. The design consists of three strip SiN waveguides with rectangular cross-section ($w=650$~nm and $t=220$ nm) forming an in-plane ``T-shaped'' arrangement \cite{EspinosaSoria2016,EspinosaSoria2018,Christiansen2020}: one waveguide supplies the excitation signal and the other two, symmetrically placed facing to each other, collect the output Stokes signals. It should be noted that SiN provides a number of practical advantages, such as long propagation distances beyond the cm-scale, transparency from visible to telecom wavelengths, cost-efficient and mass-volume production, repeatability, and CMOS compatibility \cite{RomeroGarcia2013,Subramanian2015}. Yet, for our purposes, the main feature of SiN is the operational bandwidth for guiding light, spanning the relevant visible and near-infrared (NIR) spectral range at which the fundamental plasmonic resonance (characterizing the fundamental \textit{quasinormal~mode}) of the NPoM occurs \cite{Kongsuwan2020,Ciraci2012,Baumberg2019}. The enhancing NPoM structure itself is realized in the gap region in between the three waveguides. Fitted to this crossing region, a metallic gold layer (acting as a mirror) is placed over a silica (SiO$_2$) substrate, whose thicknesses are fixed to $h_{\rm mirror}=20$ nm and $h_{\rm substrate}=55$ nm, respectively. Finally, in order to achieve the extreme light confinement steered by the excitation of the plasmonic cavity modes, we consider a spherical gold NP above the mirror \cite{Mock2008,Baumberg2019,Ciraci2012,Kongsuwan2020,Chikkaraddy2016,Benz2016a,Carnegie2018,Hill2010,Benz2016b,Chikkaraddy2017,Kongsuwan2018}, separated from each other by a self-assembled monolayer (SAM) of thickness $g_{\rm SAM}$ and refractive index set to $n_{\rm SAM}=1.5$. Both the NP radius and the gap thickness are the parameters to control the spectral behavior of the system, keeping the overall geometry with the values set above. It is worth noting that, in contrast to most previous approaches for the realization of on-chip SERS \cite{Cao2018,Zhao2018,Peyskens2016,Dhakal2016a,Raza2019,Peyskens2018}, in which all the events (namely, the excitation, enhancement, and readout of the Raman signal) were mediated by the evanescent tail of the guided mode surrounding the waveguide core \cite{Dhakal2014,Dhakal2015}, this embedded arrangement brings about a higher performance \cite{Losada2019}, increasing the overlap between the guided field and the structure \cite{EspinosaSoria2016}, and thus improving further the excitation and readout efficiency \cite{EspinosaSoria2018}. In addition to the above beneficial considerations, this proposal is further intended to provide a simplification of the illumination scheme in comparison with the free-space case \cite{Turk2019}, specifically, by removing the necessity of using tilted-angle illumination and collection for the fundamental NPoM mode \cite{Kongsuwan2020}.

\begin{figure}[t!]
	\centering
	\includegraphics[width=0.9\linewidth]{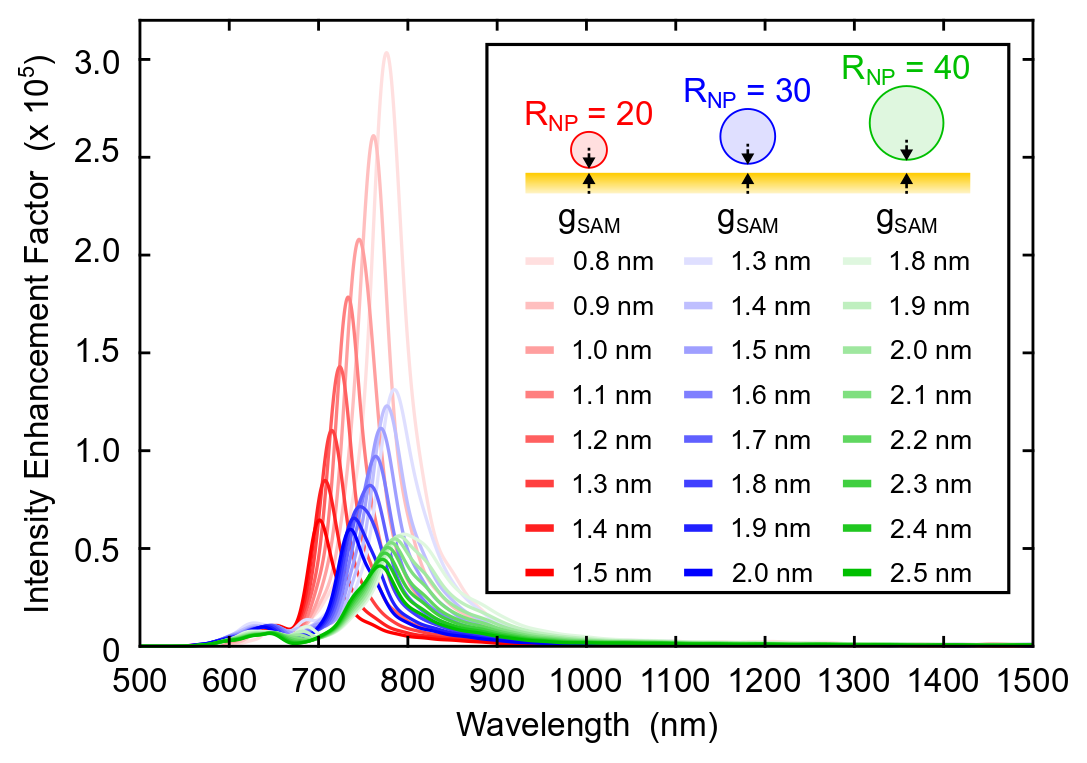}
	\caption{Intensity enhancement factor for different NP's radii and gap sizes, as indicated in the inset legend. The NPoM is excited by the fundamental TM mode ($y$-polarized) of the waveguide.}
	\label{fig02}
\end{figure}

From an optical viewpoint, SERS can be simply understood as a two-step process \cite{Ding2017}, involving the enhanced excitation of the target molecules, and the spontaneous emission of the wavelength-shifted (anti-)Stokes signal, as a result of the inelastic Raman scattering \cite{Novotny,Maier}. Based on an analytical model for on-chip SERS \cite{Peyskens2016}, the yield of such events can be characterized in terms of the \textit{Raman enhancement factor} (${\rm EF}_{\rm Raman}$), and the coupling strength ($\beta$-factor) of the SERS signal through the output waveguides. Leaving aside any short-range contribution of chemical nature \cite{Kneipp1997,Thomas2013}, at a given location (i.e., in the hotspot region where the molecules are assumed to be located, as shown in Figure \hyperref[fig01]{1c}), ${\rm EF}_{\rm Raman}({\bf r},\lambda_{\rm P},\lambda_{\rm S})={\rm EF}({\bf r},\lambda_{\rm P})^2\cdot{\rm EF}({\bf r},\lambda_{\rm S})^2$, with ${\rm EF}({\bf r},\lambda)$ standing for the local field enhancement at the pump~($\lambda_{\rm P}$) and Stokes ($\lambda_{\rm S}$) wavelengths. In turn, ${\rm EF}({\bf r},\lambda)\equiv \left|{\bf E}({\bf r},\lambda)\right|/\left|{\bf E}_0({\bf r},\lambda)\right|$, where ${\bf E}$ and ${\bf E}_0$ are the local electric field densities in the middle of the SAM's gap, with and without NPoM structure, respectively. By assuming the spherical NP to be perfectly aligned with the optical axis of the three waveguides, i.e., in the central position over the mirror surface (Figure \hyperref[fig01]{1a}), the intensity enhancement factor, ${\rm EF}({\bf r},\lambda_{\rm P})^2$, for different radii and gap sizes is shown in Figure \hyperref[fig02]{2}. In all the cases, the local field profiles are computed by means of full 3D numerical simulations, performed with the aid of the commercial solver \textit{CST Microwave Studio}. Specifically, using a point-like field monitor just below the NP, we recorded the electric field components resulting from the excitation of the NPoM nanocavity by the fundamental TM guided mode (polarized along the $y$-axis) exiting the input waveguide. Here, it is important to highlight that, owing to the broadband character of the plasmonic resonances, $\lambda_{\rm P}\approx\lambda_{\rm S}$, in such a manner that the Raman enhancement factor, ${\rm EF}_{\rm Raman}$, is generally assumed to scale approximately with the fourth power of the local field enhancement ${\rm EF}({\bf r},\lambda)$ \cite{Kneipp1997,GarciaVidal1996,Novotny,Maier,LeRu2006}. Even though this is definitely not completely true for the case of NPoM configurations, to a first approximation, we might obtain ${\rm EF}_{\rm Raman}$ from the results presented in Figure \hyperref[fig02]{2} just by calculating the square of each of the corresponding~spectra.

\begin{figure}[t!]
	\centering
	\includegraphics[width=0.9\linewidth]{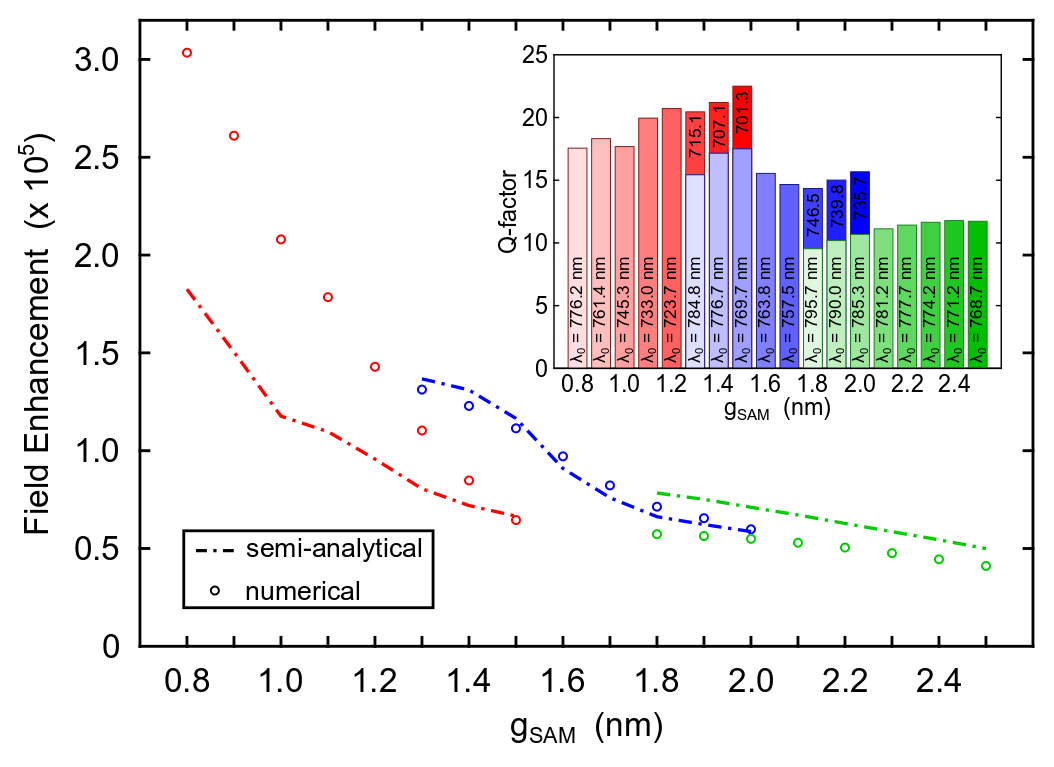}
	\caption{Field enhancement at the peak maxima for different values of NP radius and gap size (color coding as in Figure \hyperref[fig02]{2}). Circle markers represent numerical results directly obtained from Figure \hyperref[fig02]{2}, and dashed lines stand for semi-analytical curves from Eq. \eqref{FEF}, substituting the corresponding $Q$-factors (see inset barplot). Exact values of wavelength peaks ($\lambda_0$) are also indicated.}
	\label{fig03}
\end{figure}

In view of the results shown in Figure \hyperref[fig02]{2} we highlight four major features. First is about the broadband behavior, with resonant peaks covering a spectral range from $700$ nm ($1.77$~eV) to $800$ nm ($1.55$ eV), which are indeed in good agreement with theoretical predictions \cite{Kongsuwan2020}. The second remark refers to the spectral tunability of the enhancement peaks that, for a fixed NP radius ($R_{\rm NP}$), are slightly blue-shifted as the gap size ($g_{\rm SAM}$) increases; while for a given gap thickness, are strongly red-shifted as the NP radius goes up. The third observation relies on the overall trend of increasing the enhancement factor as either the NP radius increases or the gap size gets smaller. Indeed, by examining closely the plot one can realize that, for a fixed gap size the enhancement factor augmentation is directly proportional to the NP~radius. This behavior is actually well described by the following approximate formula \cite{Baumberg2019}:
\vspace{-0.05cm}
\begin{equation} \label{FEF}
{\rm EF}_{\rm peak}^2\equiv \frac{|{\bf E}_{\rm peak}|^2}{|{\bf E}_0|^2}=\left(16 \ln{2}\right)Q\cdot n_{\rm SAM}\left[\frac{R_{\rm NP}}{g_{\rm SAM}}\right]^2.
\vspace{-0.05cm}
\end{equation}
Importantly, besides matching qualitatively the aforementioned geometrical dependencies, quantitatively, numerical results also agree remarkably well. This is shown in Figure \hyperref[fig03]{3}, where we compare the field enhancement at the peak maxima obtained both numerically, that is, directly from curves shown in Figure \hyperref[fig02]{2}, and semi-analytically, i.e., according to Eq. \eqref{FEF}. Notice that, in this latter case, one has first to compute the corresponding quality factors, $Q=f_{\rm 0}/\Delta f$, somehow related with the \textit{spectral mode energy density} \cite{Maier}, where $f_{\rm 0}$ is the frequency at the peak maximum, and $\Delta f$ is given by the full-width at half-maximum (FWHM) of the intensity enhancement factor (see inset of Figure \hyperref[fig03]{3}). As can be appreciated, the best agreement between numerical and semi-analytical outcomes occurs for $R_{\rm NP}=30$ nm, displaying a fairly good matching. Finally, the last remark concerns the overall magnitude of the intensity enhancement factors attained in this particular integrated approach based on a self-assembled NPoM configuration, which largely overcomes, even by orders of magnitude (while easing the fabrication process \cite{Lee2019}), to those previously reported that make use of lithographically-defined nanoantennas with complex shapes \cite{Losada2019}.

\begin{figure}[t!]
	\centering
	\includegraphics[width=0.9\linewidth]{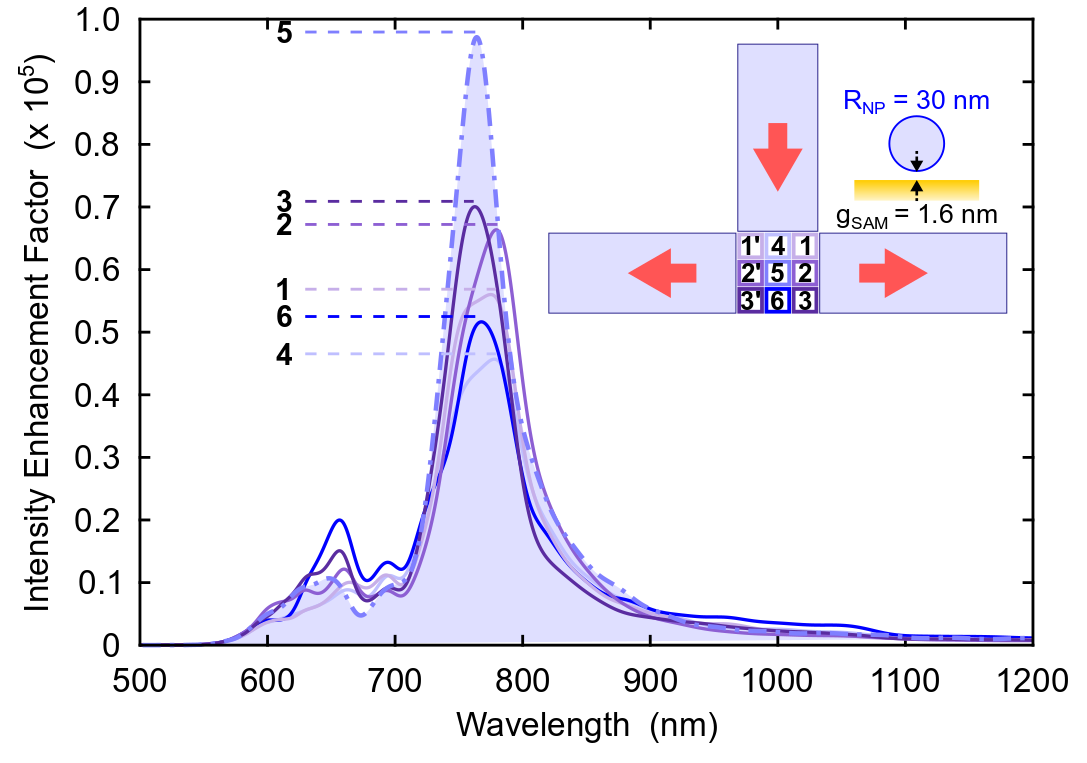}
	\caption{Intensity enhancement factor for different NP positions on the mirror plane, as indicated in the inset top-view legend. The NPoM excitation is the same as in Figure \hyperref[fig02]{2}, i.e., it is undertaken by the fundamental TM mode of the SiN waveguide.}
	\label{fig04}
\end{figure}

Obviously, the ideal case in which the NP lies just in the central position over the mirror surface is quite improbable in a realistic situation, since the metallic spheres will be randomly distributed (see, e.g., the distribution illustrated in Figure \hyperref[fig01]{1b}). In this vein, in Figure \hyperref[fig04]{4} we analyze the effect of varying the NP position over the mirror plane, considering only the particular case of $R_{\rm NP}=30$ nm and $g_{\rm SAM}=1.6$ nm. To do so, the SAM is partitioned in nine different regions as indicated in the inset (notice that, for symmetry reasons, only six partitions are meaningful). Then, assuming the NP to be located in the center of each partition, we compute numerically the local electric field density in the middle of the SAM's gap with (${\bf E}$) and without (${\bf E}_0$) the NPoM structure, following the same procedure sketched out above. Strikingly, it can be observed that the largest enhancement is achieved at the central region (i.e., partition $5$), then suggesting the involvement of nontrivial interference effects that may be attributed to the plasmonic character of the whole NPoM structure. This is therefore a straightforward way in which one may slightly modulate the enhancement factor peak at a given wavelength, still keeping the same order of magnitude.

\begin{figure}[t!]
	\centering
	\includegraphics[width=0.9\linewidth]{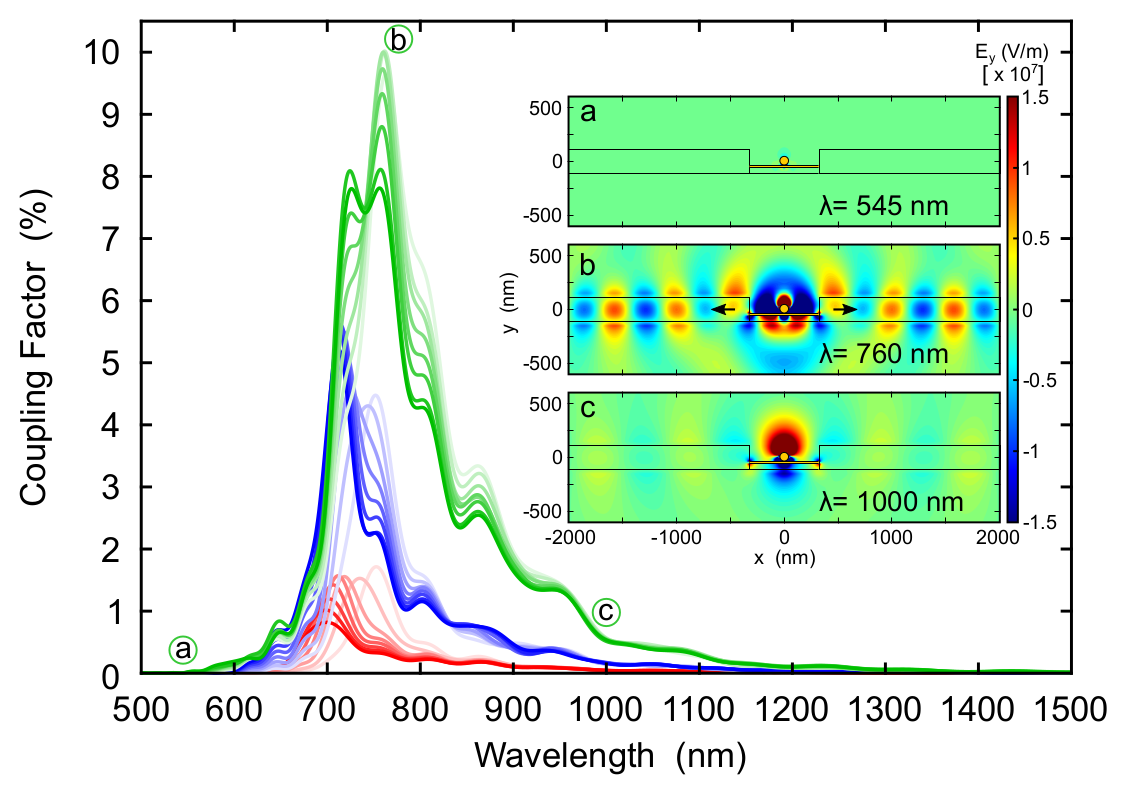}
	\caption{Coupling strength ($\beta$-factor) for different NP's radii and gap sizes. The NPoM is excited by a point-like dipolar source placed in the middle of the SAM gap just below the NP, oscillating along the $y$-direction. Inset depicts snapshots of the electric field distribution at different wavelengths, showing the largest collection efficiency in the middle~panel. The~color~scale used for different $R_{\rm NP}$ and $g_{\rm SAM}$ is the same as in Figure \hyperref[fig02]{2}.}
	\label{fig05}
\end{figure}

So far, we have only been focused on the first step of the SERS process, dealing with the enhanced excitation of the NPoM resonance by the fundamental TM mode of the SiN waveguide. On the other side, as for the Raman signal collection, we have to emulate a radiating source mimicking a Raman molecule \cite{Christiansen2020}. As a first approximation, this can be simply modeled by considering a point-like dipolar source placed in the middle of the SAM gap, just below the NP, oscillating along the $y$-direction. In this situation, the (anti-)Stokes signals are recorded at the output waveguide facets. How efficient this coupling process is will be quantified by the $\beta$-factor \cite{Peyskens2016,Dhakal2016a,Raza2019,Peyskens2018,Losada2019}, which, for our purposes, is defined as the ratio $\beta=P_{\rm TM}/P_{\rm Raman}$, where $P_{\rm TM}$ is the SERS power coupled into the fundamental TM guided mode at the output waveguides (taking into account the sum of both contributions exiting the right and left sides), and $P_{\rm Raman}$ is the total power emanating from the point-like source, i.e., the dipole acting as a sort of Raman center. These ratios are represented in Figure \hyperref[fig05]{5} for different radii and gap sizes. In addition, for the particular case of $R_{\rm NP}=40$ nm and $g_{\rm SAM}=2$ nm, we also show the electric field distribution at three distinct wavelength in order to illustrate different coupling regimes. In the light of these results, it is noteworthy to realize that, as opposed to the intensity enhancement, the overall coupling factor is larger as the radius goes up. In this respect it should be noted that, since the radiation pattern is to be identical regardless of the NP size, what actually changes is the nonradiative absorption. Be that as it may, this fact allows us to surmise the existence of a value of $R_{\rm NP}$ at which there is a trade-off performance between the enhanced Raman excitation and the collection~processes.

\begin{figure}[t!]
	\centering
	\includegraphics[width=0.9\linewidth]{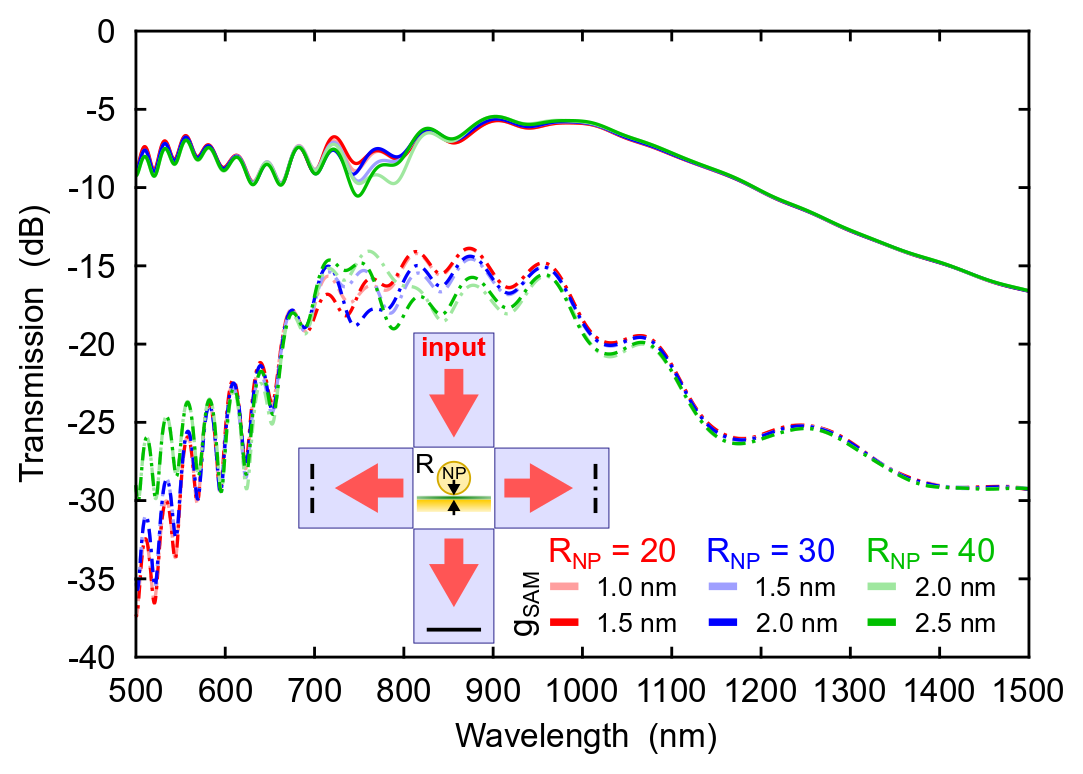}
	\caption{Numerically simulated normalized transmission spectra (in dB units) of the fundamental TM guided modes through the orthogonal (dashed) and the in-front (solid) output waveguides.}
	\label{fig06}
\end{figure}

Aiming to exploit optimally the proposed arrangement for integrated SERS spectroscopy, the signal-to-noise ratio at the Stokes wavelengths has to be maximized at the output of the chip. Therefore, from a practical point of view, it is of paramount importance not only enhancing the signal excitation and the coupling efficiency, as undertaken so far, but also minimizing the unwanted SiN Raman background collection. Indeed, it should be noted that such a background Raman signal, intrinsic to the SiN waveguide structure, can become highly detrimental \cite{Raza2019,Dhakal2017}, mainly when long (mm-scale) waveguides need to be included to drive the plasmonic cavity modes in the NPoM structure. In this sense, for our scheme it is expected that the SiN background follows the path of the incoming signal, while being highly attenuated towards the lateral directions. To analyze more in depth this reduction in our system, we consider an extra ancillary waveguide aligned with that supplying the excitation signal, together with the previous set transversely arranged (see inset of Figure~\hyperref[fig06]{6}). By simulating this structure, we observe that, independently of the specific geometrical parameters, after interacting with the NPoM, most of the output signal exits through the in-front waveguide (see Figure \hyperref[fig06]{6}), as expected, thereby ensuring that most of the Raman signal accumulated along the input waveguide would be completely suppressed through the transversal outputs. These results show that, at the plasmon resonance, about 10\% of the transmitted light from the excitation waveguide is scattered by the NPoM into the collection waveguides, with a reduction of the SiN Raman background of about $6$ dB. Since~the signal is collected in two waveguides simultaneously, the total Raman signal at the output is duplicated, and hence, the total improvement of the signal-to-noise ratio would be around $10$ dB using our ``T-shaped'' architecture. As can be seen, this outcome is rather broadband though, so likely is controlled by the geometrical considerations. Therefore,~using~waveguides~with curved endings, making a type of facet lens, would be helpful to suppress further the unwanted Raman background signal, while improving the NPoM coupling efficiency. Although currently beyond the present work, this might further reduce the noise of this background.

In conclusion, we have carried out a numerical study to assess the viability for realizing SERS in an integrated platform exploiting the NPoM structure optically-excited through SiN waveguides. Besides the practical benefits which directly stem from the miniaturized approach, the main advantages of the proposed arrangement, in contrast with the free-space case, arise both from the ease and the foreseeable reduction of costs to experimentally implement the enhanced excitation and collection of plasmonic cavity modes in NPoM configurations. Furthermore, this integrated approach presents important features for on-chip SERS spectroscopy, such as tunability (depending on geometrical parameters involving the SAM gap and the NP), or high-performance, with intensity enhancement factors beyond $10^5$. Noticeably, these quantities exceeds, in up to two orders of magnitude, with respect to earlier results using lithographically-defined nanoantennas with complex geometries \cite{Losada2019}, with the additional benefit of simplifying the fabrication process \cite{Lee2019}, as NPoM structures can be readily made by self-assembly techniques. We have also shown that Raman signals can be straightforwardly collected in transversely placed waveguides, with coupling efficiencies reaching up to 10\%. Nonetheless, this performance, as well as the signal-to-noise ratio, could likely be improved by engineering further the NPoM interface, for instance, by varying the remaining geometric parameters of the system (e.g., the height of the mirror), by investigating other SERS geometries \cite{Maier}, by regarding the presence of roughness or nanometric crevices on the mirror plane \cite{GarciaVidal1996,Ciraci2020}, or by looking into other materials with lower absorption losses (e.g., considering silver, copper \cite{Naik2013}, or even artificially engineered metamaterials \cite{Wang2019}). Moreover, it would also be desirable to look into simpler geometrical arrangements, e.g., ones based on back-scattering, which, in addition to allowing a considerable reduction of the whole footprint of the system, would further simplify the implementation of the excitation and readout of the Raman signal. This work constitutes a good starting point to approach the optical response of plasmonic cavity modes enabled by NPoMs embedded in Si-based photonic waveguides. Along with SERS spectroscopy \cite{Raza2019}, this scheme might also be exploited for other nonlinear functionalities, such as frequency mixing, all-optical switching, or modulation \cite{Leuthold2010,Kauranen2012}, in hybrid plasmonic-photonic integrated platforms \cite{Fortuno2016,EspinosaSoria2016,EspinosaSoria2018}.

\begin{acknowledgement}
We thank R. Baets, A. Raza, S. Clemmen (UGent), and R. Chikkaraddy (Cambridge) for fruitful discussions. This work was supported by funding from the European Commission Project THOR H2020-EU-829067.
\end{acknowledgement}

\section{Graphical TOC Entry}
\begin{center}
	\fbox{\parbox[][160pt][c]{260pt}{\centering\includegraphics[width=1.0\linewidth]{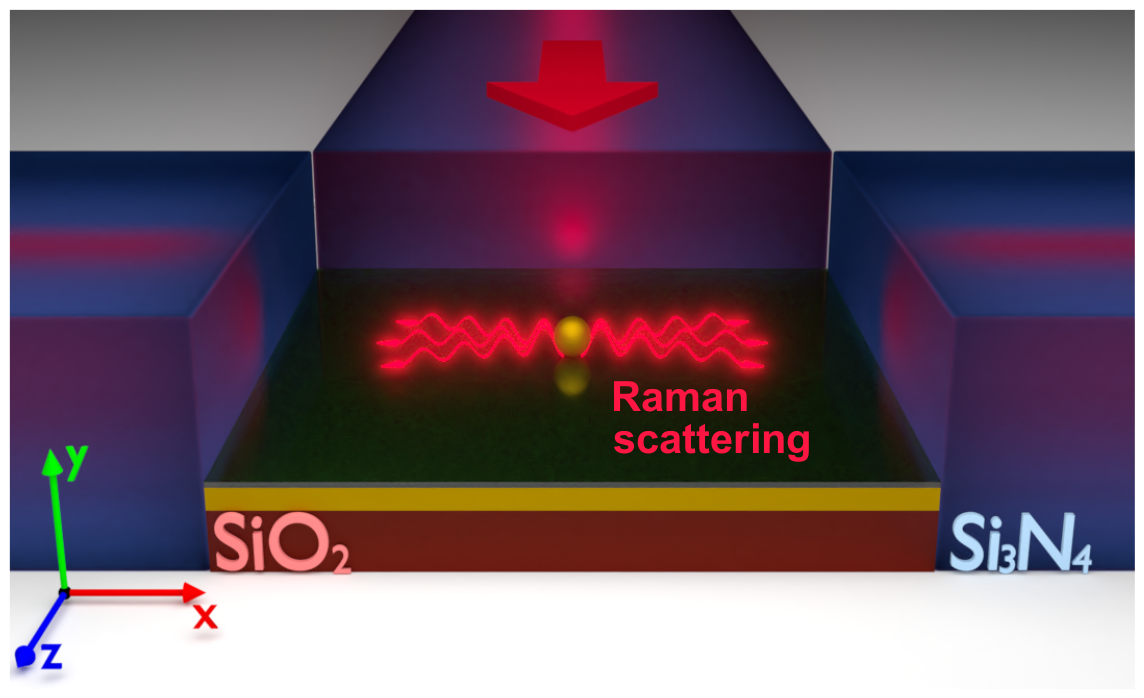}}}
\end{center}

\end{document}